\documentclass[aps,prb,reprint,superscriptaddress,showpacs,floatfix]{revtex4-1}
\usepackage{amsfonts}
\usepackage{amsmath}
\usepackage{amssymb}
\usepackage{mathbbol}
\allowdisplaybreaks[1]
\usepackage{graphicx}
\usepackage{hyperref}
\usepackage{url}
\usepackage{color}
\newcommand{\rem}[1]{}

\newcommand{\ave}[1]{ {\langle #1 \rangle} }

\newcommand{\abs}[1]{ \lvert #1 \rvert }

\begin{document}
%
\title{
 Strong electronic correlation in the Hydrogen chain: a variational Monte Carlo study
}
\author{Lorenzo Stella}
\email{lorenzo.stella@ehu.es}
\affiliation{
  Nano-Bio Spectroscopy group and ETSF Scientific Development Centre,
  Dpto. F\'{i}sica de Materiales, Universidad del Pa\'{i}s Vasco,
  Centro de F\'{i}sica de Materiales CSIC-UPV/EHU-MPC and DIPC,
  Av. Tolosa 72, E-20018 San Sebasti\'{a}n, Spain}
\affiliation{
  Department of Physics and Astronomy and London Centre for Nanotechnology, 
  University College London, Gower Street, London WC1E 6BT, United  Kingdom}
\author{Claudio Attaccalite}
\affiliation{
  Institut N\'{e}el, CNRS/UJF 25 rue des Martyrs BP 166, B\^{a}timent D 38042 Grenoble cedex 9 France}	
\author{Sandro Sorella}
\affiliation{
  SISSA, Via Bonomea 265, 34136 Trieste, Italy and Dipartimento di Fisica, Universit\`{a} 
  di Trieste, strada Costiera 11, 34151 Trieste, Italy}	
\author{Angel Rubio}
\affiliation{
  Nano-Bio Spectroscopy group and ETSF Scientific Development Centre,
  Dpto. F\'{i}sica de Materiales, Universidad del Pa\'{i}s Vasco,
  Centro de F\'{i}sica de Materiales CSIC-UPV/EHU-MPC and DIPC,
  Av. Tolosa 72, E-20018 San Sebasti\'{a}n, Spain}	
\affiliation{     
  Fritz-Haber-Institut der Max-Planck-Gesellschaft, Berlin, Germany}	
\begin{abstract}
In this article, we report a fully \emph{ab initio} 
variational Monte Carlo study of the linear, and periodic chain of Hydrogen atoms,
a prototype system providing the simplest example of strong electronic correlation 
in low dimensions.
In particular, we prove that numerical accuracy comparable to that of benchmark 
Density Matrix Renormalization Group calculations can be achieved
by using a highly correlated Jastrow-antisymmetrized geminal power  
variational wave function.
Furthermore, by using the so-called ``modern theory of polarization''
and by studying the spin-spin and dimer-dimer correlations functions,
we have characterized in details the crossover between the weakly 
and strongly correlated regimes of this atomic chain.
Our results show that variational Monte Carlo provides 
an accurate and flexible alternative to highly correlated methods 
of quantum chemistry which, at variance with these methods, 
can be also applied to a strongly correlated solid 
in low dimensions close to a crossover or a phase transition.
\end{abstract}
\maketitle
%
\section{Introduction}\label{intro}
%
The homogeneous (i.e., equispaced), linear, 
and periodic chain of Hydrogen atoms (hereafter, the H-chain) is
commonly believed to be the simplest physical system described
by the one-band, periodic, one-dimensional (1D) Hubbard Hamiltonian
\cite{Hubbard64,*Gebhard_book}
(see Eq.~\ref{hubbard:eqn}).
This Hamiltonian is exactly solvable
\cite{Lieb68,*Lieb03}
and its solution predicts the H-chain to be always a Mott-Hubbard 
(i.e., a strongly correlated) insulator.
Indeed, it seems reasonable to model the H-chain by including only the $1s$
orbitals and by neglecting the long-range tail of the Coulomb interaction, 
especially for large interatomic distances.
As a consequence, the H-chain has been intensively studied to benchmark 
\emph{ab initio} approaches to strong electronic correlation
\cite{Hachmann06,Tsuchimochi09,Lin11,Sinitskiy10,*Mazziotti11}  
despite this atomic chain is not directly observable
due to the well known Peierls' instability.

Previous \emph{ab initio} studies---mostly using methods of
quantum chemistry---have been focused on finite (i.e., not periodic)
linear chains, only.
However, Periodic Boundary Conditions (PBC) analogous to the Born-von K\'{a}rm\'{a}n 
boundary conditions used in solid state physics are better suited to investigate
phase transitions or crossovers. 
Indeed, unwanted edge effects are avoided by using PBC
and a speed up of the convergence to the thermodynamic limit, 
i.e., the limit of infinite linear chains, is expected. 
\cite{Resta99,Capello05a}
Hence, an \emph{ab initio} description 
of the low energy physics of a properly periodic H-chain
is still missing.
In this article we provide the first exhaustive, 
fully \emph{ab initio}
\footnote{
We compute the electronic ground state of the full non-relativistic 
Hamiltonian with fixed atomic geometry, but without using pseudopotentials.
}	
variational description of periodic chains by using the same kind of variational 
wave function for both the weakly and the strongly correlated regimes, 
i.e., for both small and large interatomic distances.

From previous studies, it is known that
in the strongly correlated regime---i.e., for interatomic distances, $a$, larger
than a certain critical distance, $a_c$---the ground state of the finite H-chains
is characterized by a huge degeneracy of the natural orbitals
\cite{Hachmann06,Tsuchimochi09}
which leads to an almost uniform natural orbital population,
narrowly dispersed around $1$.
\cite{Sinitskiy10,*Mazziotti11}
This behavior has to be contrasted with the weakly correlated regime 
($a<a_c$) for which
already the restricted Hartree-Fock reference---that yields 
either doubly occupied of empty 
natural orbital---is quite accurate.
\cite{Karpfen79,*Kertesz79}

In many cases, 
\cite{Tsuchimochi09}
part of the static correlation which characterizes the strongly correlated 
regime can be effectively recovered by means of an unrestricted Hartree-Fock 
calculation, 
or equivalently, by means of a spin-polarized density functional theory calculation within the
local density approximation.
\cite{Gebhard_book}
Although justified for finite systems, a spin-polarized approach implies a mean-field 
antiferromagnetic order, which cannot be trusted in the thermodynamic limit, because 
true antiferromagnetism is not possible for 1D solids.
\cite{Affleck89}

The Density Matrix Renormalization Group (DMRG) provides an, in principle 
exact, algorithm to compute the electronic structure of 1D and
almost-1D systems, although in practice limited by the size of the
orbital basis set.
\cite{Chan11}
It works very efficiently also when other highly correlated methods fail, 
e.g., configuration interaction
\cite{Cimiraglia81,*Vetere08}
is not applicable if the system size is too large and
the standard coupled cluster singles and doubles plus perturbative triples
becomes unstable in 1D for large interatomic distances.
\cite{Hachmann06}
Nevertheless, even the very favorable numerical efficiency of DMRG is lost 
for a gapless (i.e., metallic) chain.  
In this case, also the M\/{o}ller-Plesset perturbative
approach is not straightforwardly applicable due to the numerical issues 
triggered by the vanishing small gap.
%
\section{Computational methods}\label{methods}
%
Among the possible 
alternatives to standard quantum chemical approaches,
\cite{Tsuchimochi09,Sinitskiy10,*Mazziotti11}
one can consider non-perturbative
quantum Monte Carlo (QMC) 
methods like VMC or the, in principle
more accurate, Diffusion Monte Carlo (DMC).
\cite{Foulkes01}            
In fact, a direct application of DMC to a homogeneous chain 
raises some ergodicity issues when the electrons are
very localized about the nuclei, i.e., in the
strongly correlated regime.
\cite{Casula05,*Casula06}
As a consequence, to date only alternating (i.e., dimerized) 
chains have been investigated by means of DMC
\cite{Umari05,*Umari09}
On the other hand, VMC can be made ergodic by tailoring 
the sampling and by improving the variational many-body wave function, 
so that it remains very effective also close to crossovers
and phase transitions.
\cite{Resta99,Capello05a}

There have been dramatic theoretical advances in the quality of the variational 
wave function during the last decade, so that it is now possible to achieve chemical 
accuracy for atoms, ions, small molecules and even periodic systems.
\cite{Drummond04,*Seth11,Casula04,Marchi09,*Azadi10,Sorella07,Sorella11}
These quantitative improvements have been made possible 
by new stochastic optimization techniques
which can optimize variational wave functions depending on hundreds free parameters.
\cite{Umrigar06}

In our variational calculations, we have used 
a Jastrow-antisymmetrized geminal power (JAGP) variational wave function
for $N$ (even) interacting electrons,
\cite{Casula04,Marchi09,*Azadi10}
\begin{equation}\label{JAGP} 
  \Psi_{N}(\vec R) = J(\vec R) {\cal A } \prod
  _{i=1}^{N/2} \Phi (\vec{r}^\uparrow_i, \vec{r}^\downarrow_i ) \;,
\end{equation}
where $\vec R= \left\{ \vec r_1^\uparrow, 
\dots, \vec r_{N/2}^\uparrow, r_1^\downarrow, \dots ,  
\vec r_{N/2}^\downarrow \right\}$  
is the $3N$-dimensional coordinate vector,
${\cal A}$ is the antisymmetrization operator, 
$\Phi (\vec{x}^\uparrow,\vec{y}^\downarrow)$ is
a symmetric function describing a singlet electron pair
and
\begin{equation}\label{jastrow} 
  J(\vec R) = \exp \sum_{i,j}^N [u( r_{i,j} )  
  + f(\vec r_i, \vec r_j )]\;,
\end{equation}
is the Jastrow factor.

By using Eq.~\ref{JAGP}, one can accurately 
describe both static and dynamics correlations.
Indeed, the antisymmetrized geminal power, 
${\cal A } \prod_{i=1}^{N/2} \Phi (\vec r^\uparrow_i, \vec r^\downarrow_i )$
provides a very compact multideterminant reference,
while the Jastrow factor
takes into account the dynamic correlation from: 
i) A short-range  homogeneous electron-electron interaction
through the term $u( r_{i,j})$ which just
depends on the distance, $ r_{i,j}= | \vec r_i - \vec r_j |$, between paired electrons;
\cite{Foulkes01} 
ii) The inhomogeneous electron-electron-nucleus and electron-electron-nucleus-nucleus interactions
through the term $f(\vec r_i, \vec r_j )$,
which depends separately on the coordinates of the paired electrons, and 
it can also describe long-range electronic correlations.

Results showed in this article have been obtained using
a short-range homogeneous Jastrow factor, $u(r_{i,j})=(b/2)(1-e^{-b r_{i,j}})$.
In addition, to asses the sensitivity of these results on the detail of the 
wave function, we have also considered a long-range homogeneous Jastrow factor,
$u(r_{i,j})=r_{i,j}/2(1+b r_{i,j})$ (see Fig.~\ref{fig_z_a5.0_convergence:fig}).
In both cases, $b$ is a variational parameter, and the electronic cusp conditions
\cite{Foulkes01}
are automatically satisfied.

The functions $\Phi (\vec{x}^\uparrow,\vec{y}^\downarrow)$ in Eq.~\ref{JAGP} 
and $f(\vec x, \vec y )$ in Eq.~\ref{jastrow} are expanded 
using (in principle different) nonorthogonal atomic orbitals,
\cite{Petruzielo10}
$\{\phi_i\}$ and $\{\varphi_i\}$ so that
\begin{eqnarray}
\Phi( \vec{x}^\uparrow,\vec{y}^\downarrow)&=&\sum_{\alpha,\beta}\lambda_{\alpha,\beta}\varphi_{\alpha}
(\vec{x}^\uparrow)\varphi_{\beta}(\vec{y}^\downarrow)\;, \nonumber \\
f(\vec{x},\vec{y})&=&\sum_{\alpha,\beta} g_{\alpha,\beta}\phi_{\alpha}
(\vec{x})\phi_{\beta}(\vec{y})\;.
\label{jastroweq}
\end{eqnarray}
In particular, up to $3s$ orbitals for the geminal part and $2s2p$ for the Jastrow
part have been considered in this work.
\footnote{
 Since expectation values of the variational wave function
 are not computed analytically, but stochasticly
 \protect\cite{Foulkes01}, both Gaussian type orbitals
 or Slater type orbitals can be used with the same efficiency.
}	

In principle, all the entries of the matrices $\lambda_{\alpha,\beta}$ and
$g_{\alpha,\beta}$ in Eq.~\ref{jastroweq} are variational parameters
to be optimized.
However, by taking advantage of the symmetries of the periodic linear chain,
\cite{Sorella11}
and by using an alternative, minimal expansion in terms of 
molecular orbitals,
\cite{Marchi09,*Azadi10}
the actual number of independent variational parameters to optimize is reduced,
so that the optimization can be effectively performed by the method
describes in Ref.~\onlinecite{Umrigar06}.

All the VMC calculations have been performed by using the {\it TurboRVB}
package,\cite{turborvb} starting from a density functional theory 
(in local density approximation) calculation  
employing the orbital basis set, $\{\phi_i\}$, of the geminal part
(see Eq.~\ref{jastroweq}).
This preliminary step is done to speed up the convergence of the following VMC 
optimization, while avoiding an uncontrolled bias.

Finally, we have used a supercell with PBC 
\cite{Foulkes01}  
in all three Cartesian directions
to model the periodic linear chain.
To avoid unphysical self-interaction of the chain with its 
periodic replicas, the transverse dimensions of the supercell 
have been taken as ${\rm min}(16a,80)$ a.u., where $a$ is the 
interatomic distance.
According to the supercell formalism, the thermodynamic limit, 
$N \to \infty$, can be achieved by increasing the number, $N$, 
of H atoms in the supercell.
In particular, where not otherwise indicated, by H$_N$ we mean
a periodic chain with $N$ atoms in the supercell.

To identify the weakly and the strongly correlated regimes,
we have used the so-called ``modern theory of polarization''.
\cite{Kohn64,*Resta94,*Hine07}
This theory also provides a way to discriminate between a metal 
and an insulator alternative to the knowledge of the (many-body) charge gap,
which in fact is not accessible by a variational ground state calculation.

In practice, by VMC one computes the expectation values of the 
complex polarization, $z_N$,
\cite{Umari05,*Umari09}
\begin{equation}\label{zeta}
  z_N = \langle\Psi_N|e^{i\frac{2\pi}{L}\sum_i r_i^{\parallel}}|\Psi_N\rangle\;,\\
\end{equation}
where $ r_i^{\parallel} $ is the component of $ \vec r_i $ parallel to
the chain axis.
Then the electronic localization length, $\lambda_N $, is obtained as
\begin{equation}\label{localization}
  \lambda_N = \left(\frac{L}{2 \pi} \right ) \sqrt{-\frac{\ln |z_N|^2}{N}}\;,
\end{equation}
where $L$ is the longitudinal dimension of the supercell and $N$ the number of H atoms
in the supercell.

From previous studies,
\cite{Hachmann06,Tsuchimochi09,Sinitskiy10,*Mazziotti11}
one expects a huge degeneracy of the natural orbitals
when the electrons are very localized about the nuclei,
i.e., when $\lambda_N /a \ll 1$. 
Besides, the theory says that, in the thermodynamic limit, $N \to \infty$, 
a metal is characterized by a vanishing modulus of the complex polarization, 
$\abs{z_N} \to 0$, while in the insulating case
$\abs{z_N} \to 1$.
\cite{Resta99,Capello05a}
%
\section{Results}\label{results}
%
\begin{figure}[!ht]
\begin{center}
\includegraphics[width=8.0cm]{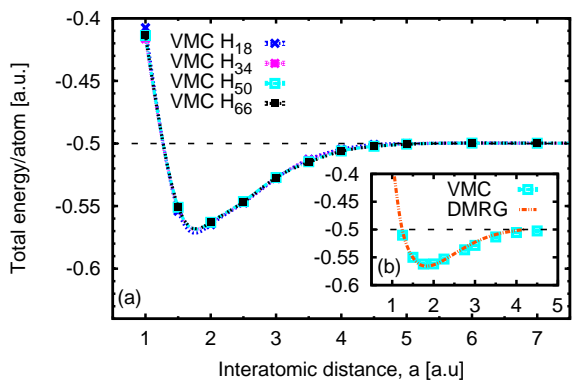}
\end{center}
\caption{
(Color online) Panel (a): Total energy per atom as a function of the interatomic distance
from VMC calculations of periodic chains with $18$, $34$, $50$, 
and $66$ H atoms in the supercell.
[Data are almost superimposed at the scale of this figure, 
see also Tab.~\protect\ref{tab_etot:tab}.]
Inset (b): Comparison between the total energy per atom of a \emph{finite} H$_{50}$ chain obtained by VMC
 and DMRG
\protect\cite{Hachmann06}.
}
\label{fig_etot:fig}
\end{figure}
In Fig.~\ref{fig_etot:fig}(a), we show the convergence 
of the total energy per atom
by increasing the number of H atoms per supercell
for several interatomic distances.
We note that the H$_{50}$ periodic H-chain is already well converged at 
the scale of this figure.
To follow the fine detail of the convergence,
the values of the total energy per atom details have been also 
listed in Tab.~\ref{tab_etot:tab}.
\begin{table}
\begin{center}
\begin{tabular}{ccccc}
\hline 
\hline 
\hspace{0.2cm} $a$ \hspace{0.2cm} & \hspace{0.5cm} H$_{18}$ \hspace{0.5cm} & \hspace{0.5cm} H$_{34}$ \hspace{0.5cm} & \hspace{0.5cm} H$_{50}$ \hspace{0.5cm} & \hspace{0.5cm} H$_{66}$ \hspace{0.5cm} \\
\hline
1.0 & -0.40751(4) & -0.41639(3) & -0.41380(3) & -0.41358(2)\\
1.5 & -0.55402(2) & -0.55156(1) & -0.55099(1) & -0.55070(1)\\
2.0 & -0.56480(2) & -0.56329(1) & -0.56296(1) & -0.56284(1)\\
2.5 & -0.54747(2) & -0.54699(1) & -0.54639(1) & -0.54682(1)\\
3.0 & -0.52796(2) & -0.52770(2) & -0.52717(1) & -0.52727(1)\\
3.5 & -0.51263(3) & -0.51308(2) & -0.51459(2) & -0.51508(1)\\
4.0 & -0.50458(3) & -0.50556(4) & -0.50599(2) & -0.50626(1)\\
4.5 & -0.50080(3) & -0.50206(1) & -0.50222(1) & -0.50237(1)\\
5.0 & -0.50014(2) & -0.50029(1) & -0.50047(1) & -0.50063(1)\\
6.0 & -0.49962(1) & -0.49971(1) & -0.49972(1) & -0.49965(1)\\
7.0 & -0.49980(1) & -0.49981(1) & -0.49979(1) & -0.49972(1)\\
\hline
\hline
\end{tabular}
\end{center}
\caption{
Total energy per atom as a function of the interatomic distance, 
$a$, for the same periodic chains of Fig.~\protect\ref{fig_etot:fig}(a).
The VMC error on the last digit is indicated in parenthesis.}
\label{tab_etot:tab}
\end{table}

In Fig.~\ref{fig_etot:fig}(b), a direct comparison
between the VMC total energy for the H$_{50}$ \emph{finite} chain
and the benchmark DMRG results obtained by using a STO-6G basis set
\cite{Hachmann06}
demonstrates the accuracy of our optimized 
JAGP variational wave function.
\footnote{
  The original results from Table VI of Ref.~\protect\onlinecite{Hachmann06} 
  have been smoothly interpolated and uniformly shifted to obtain 
  the reference of Fig.~\protect\ref{fig_etot:fig}(a).
}
In this case, to have a fair comparison against the DMRG data, PBC have not
been employed to obtain the VMC results showed in Fig.~\ref{fig_etot:fig}(b).
The difference between the total energy of H$_{50}$ chains with and without PBC and
the same interatomic distance is of the order of few mHa per atom.

\begin{figure}[!ht]
\begin{center}
\includegraphics[width=8.0cm]{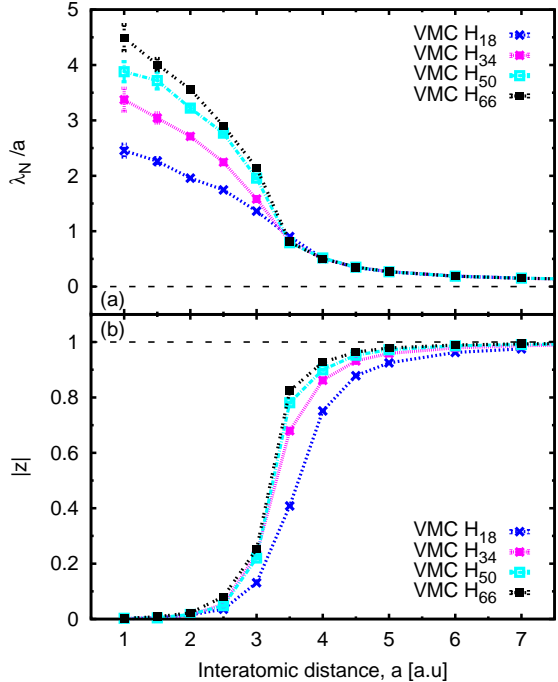}
\end{center}
\caption{
(Color online) Panel (a): Electronic localization length, $\lambda_N$, divided by the interatomic distance, $a$,
as a function of $a$, for the same chains of Fig.~\protect\ref{fig_etot:fig}(a).
Panel (d): modulus of the complex polarization, $\abs{z_N}$
as a function of the interatomic distance for the same chains of Panel (c). 
}
\label{fig_z:fig}
\end{figure}
Having verified the quality of the variational wave function,
in Fig.~\ref{fig_z:fig}(a)
we plot the electronic localization length, $\lambda_N$, in units of
the interatomic distance, $a$, as a function of $a$.
For all the supercells considered, we find that
\begin{equation}\label{critical_behavior:eqn}
 \lambda_N/a \propto \left\{
 \begin{array}{cc}
  \abs{a-a_c}^{\eta} & {\rm if} \quad a<a_c \; \\
  a^{-1} & {\rm if} \quad a>a_c \; 
 \end{array}
 \right.
\end{equation}
where $\eta \simeq 0.5$ and $a_c \simeq 3.5$ a.u.
This critical behavior is also in agreement 
with the sudden switch from $\abs{z}\simeq 0$ to 
$\abs{z}\simeq 1$ visible in Fig.~\ref{fig_z:fig}(b),
i.e., to the crossover between a (finite-size) 
metal and an insulator, namely a Mott-Hubbard insulator.
\cite{Hubbard64,*Gebhard_book}

\begin{figure}[!ht]
\begin{center}
\includegraphics[width=8.0cm]{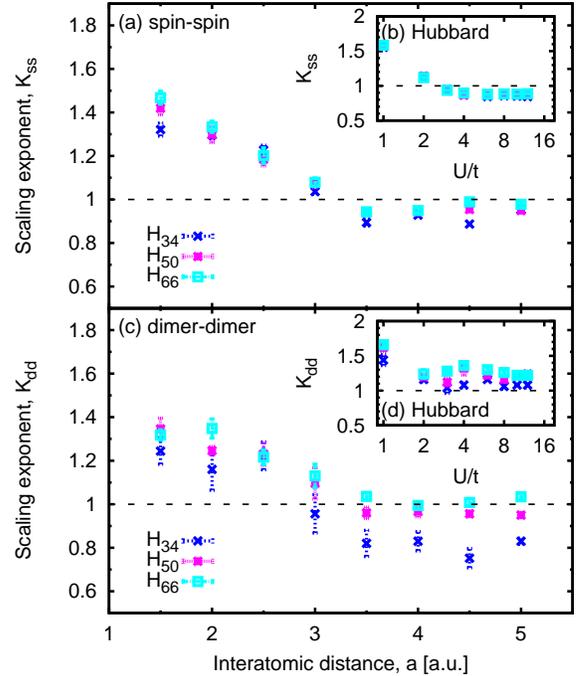}
\end{center}
\caption{
(Color online) Panel (a): Scaling exponent $K_{ss}$ of the spin-spin 
correlation function (see Eq.~\ref{Kss}) as a function of the interatomic distance,
for same H-chains of Fig.~\protect\ref{fig_etot:fig} and Fig.~\protect\ref{fig_z:fig}.
Panel (b): Scaling exponent $K_{ss}$ of the Hubbard model with a corresponding number
of sites (see text).
[Data are superimposed at the scale of this figure.]
Panel (c) and (d): same as panel (a) and (b), but for the dimer-dimer correlation function
(see Eq.~\ref{Kdd}).
}
\label{fig_expo}
\end{figure}
To further characterize the nature of the weakly and strongly correlated regimes
of the H-chain, we have investigated the spin-spin,
\begin{equation}	
f_{ss}(i-j)=\ave{\Psi_N \abs{\hat{S}_z^{(i)}\hat{S}_z^{(j)}} \Psi_N}
\end{equation}
and the dimer-dimer
\begin{equation}
f_{dd}(i-j)=\ave{\Psi_N \abs{\hat{S}_z^{(i)}\hat{S}_z^{(i+1)}\hat{S}_z^{(j)}\hat{S}_z^{(j+1)}} \Psi_N}
\end{equation}
correlation functions,
where $\hat{S}_z^{(i)}$ measures the transverse component of the electronic spin about the $i$th
H atom of the chain.
By neglecting logarithmic corrections, we have fitted these functions by
\cite{Schulz90,*Fabrizio96,*Capello05b}
\begin{eqnarray}
  f_{ss}(i-j) &=& \frac{a_{ss}}{(i-j)^2} +b_{ss}\frac{\cos(\pi (i-j))}{(i-j)^{K_{ss}}} \; \label{Kss}\\
  f_{dd}(i-j) &=& a_{dd} +b_{dd}\frac{\cos(\pi (i-j))}{(i-j)^{K_{dd}}} \;. \label{Kdd}\
\end{eqnarray}
for the spin-spin and the dimer-dimer cases, respectively.
The parameters, $a_{ss}$, $b_{ss}$, $K_{ss}$, $a_{dd}$, $b_{dd}$, and $K_{dd}$
have been fitted independently for each value of the interatomic distance, $a$, 
and the number of H atoms in the supercell.

Results for the scaling exponents, $K_{ss}$ and $K_{ss}$, of the H-chain are 
reported in Fig.~\ref{fig_expo}(a)
and Fig.~\ref{fig_expo}(c), respectively.
For comparison, in Fig.~\ref{fig_expo}(b) and Fig.~\ref{fig_expo}(d) 
we show the scaling exponents obtained by solving numerically
\cite{Calandra98}
the $N$-site Hubbard model Hamiltonian 
\cite{Hubbard64,*Gebhard_book}
with PBC (i.e., the simulation cell is folded so that the $(j\!\!+\!\!N)$th and $j$th sites 
represent the same atom)
\begin{equation}\label{hubbard:eqn}
H = -t\sum_{j=1}^N\sum_{\sigma={\uparrow,\downarrow}} (c^{\dagger}_{j,\sigma} c^{}_{j+1,\sigma} 
+c^{\dagger}_{j+1,\sigma} c^{}_{j,\sigma} ) +U\sum_{j=1}^{N} n_{j,\uparrow}n_{j,\downarrow}\;,
\end{equation}
with a number of sites, $N$, correspondent to the number 
of H atoms in the chain supercell.
In Eq.~\ref{hubbard:eqn}, the creation(annihilation) operator $c^{\dagger}_{j,\sigma}$($c^{}_{j,\sigma}$) 
creates(annihilates) an electron of spin $\sigma$ at site $j$, while
$n_{j,\sigma}=c^{\dagger}_{j,\sigma}c^{}_{j,\sigma}$.
Since one expects (for $U$ fixed) that $\log(U/t) \propto a$, we have shown the Hubbard 
exponents as a function of $U/t$ using a semi-log plot.

The behavior of the scaling exponent $K_{ss}$ is very similar in the two cases, i.e.,
H-chain in Fig.~\ref{fig_expo}(a) and Hubbard model in  Fig.~\ref{fig_expo}(b).
However, small but noticeable discrepancies between the H-chain and the Hubbard model
for the scaling exponent $K_{dd}$ are found by comparing Fig.~\ref{fig_expo}(c) 
and Fig.~\ref{fig_expo}(d).
 
For large interatomic distance $a$, both the H-chain exponents 
behave as expected for the Hubbard model
(neglecting logarithmic corrections)
i.e., $K_{dd} \sim K_{ss} \sim 1$.
\cite{Schulz90,Fabrizio96,*Capello05b}
Less conclusive results can be inferred from the weakly correlated regime 
for $a<a_c$.
Finite-size effects are responsible for the deviation of the scaling 
exponents from their thermodynamic values in the case of the Hubbard model.
The same effects also mask possible discrepancies in the thermodynamic limit 
of the H-chain and the Hubbard model.
Further numerical investigations are needed to provide conclusive 
results on a possible metal-insulator transition at a finite value, 
$a_c$, of the interatomic distance of the H-chain.
(see Sec.~\ref{conclusions}).

\begin{figure}[!ht]
\begin{center}
\includegraphics[width=8.0cm]{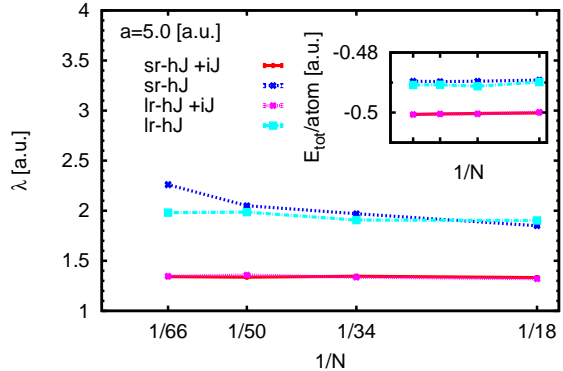}
\end{center}
\caption{
(Color online) Panel (a): Localization length as a function of the inverse of the number of
H atoms in the supercell for different parametrizations of the
JAGP variational wave function (see text).
[Data `sr-hJ+iJ' and `lr-hJ+iJ' are superimposed at the scale of this figure,
see also Tab.~\protect\ref{tab_etot_a5.0:tab}.]
Inset (b), the corresponding total energy per atom.
}
\label{fig_z_a5.0_convergence:fig}
\end{figure}
Finally, we investigate in more detail the capability of the variational JAGP wave function
to describe the Mott-Hubbard insulating phase of the H-chain for $a>a_c$.
In particular, we focus on the $a=5.0$ a.u. case and we consider some variants of 
the JAGP variational wave function, Eq.~\ref{JAGP}.

In Fig.~\ref{fig_z_a5.0_convergence:fig}(a) we plot the electronic localization length,
$\lambda_N$, obtained by optimizing different JAGP variational wave functions.
We consider the following cases (for the notation, see previous Section):
i) sr-hJ+iJ, corresponding
to Eq.~\ref{JAGP} with short-range homogeneous Jastrow factor
plus the inhomogeneous part.
This is the standard case considered elsewhere in this article;
ii) lr-hJ+iJ, as the sr-hJ+iJ wave function, 
but with a long-range homogeneous Jastrow factor, instead;
iii) sr-hJ, and iv) lr-hJ, which are as the sr-hJ+iJ and lr-hJ+iJ wave functions,
respectively, but without the inhomogeneous Jastrow factor, i.e.,
with $f( \vec r_i, \vec r_j)=0$.

We find that the localization length, $\lambda_N$, is well converged
at the scale of Fig.~\ref{fig_z_a5.0_convergence:fig}(a)
if the inhomogeneous Jastrow factor is included, regardless of
the choice of the homogeneous part.
If this factor is not included, values of $\lambda_N$
almost twice as large are found and they slightly increase with the system size,
showing that the homogeneous Jastrow alone is not enough 
to give an accurate description of the Mott-Hubbard insulating phase.
Values of the total energy per atom reported
in Fig.~\ref{fig_z_a5.0_convergence:fig}(b) also confirm the relevance of the
long-range, inhomogeneous Jastrow factor to improve the accuracy of the VMC 
description of a Mott-Hubbard insulator.
To follow the fine detail of the convergence,
the values of the total energy per atom have been also 
listed in Tab.~\ref{tab_etot_a5.0:tab}.
\begin{table}
\begin{center}
\begin{tabular}{ccccc}
\hline 
\hline 
\hspace{0.2cm} $N$ \hspace{0.2cm} & \hspace{0.2cm} sr-hJ+iJ \hspace{0.2cm} & \hspace{0.2cm} lr-hJ+iJ \hspace{0.2cm} & \hspace{0.35cm} sr-hJ \hspace{0.35cm} & \hspace{0.5cm} lr-hJ \hspace{0.35cm} \\
\hline
18 & -0.50014(2) & -0.49987(2) & -0.48918(6) & -0.48984(6)\\
34 & -0.50029(1) & -0.50042(1) & -0.48956(4) & -0.49123(3)\\
50 & -0.50047(1) & -0.50045(1) & -0.48969(3) & -0.49072(3)\\
66 & -0.50063(1) & -0.50066(1) & -0.48956(3) & -0.49082(3)\\ 
\hline
\hline
\end{tabular}
\end{center}
\caption{Total energy per atom as a function of the number, $N$, of H atoms in
the supercell, for the same chains of Fig.~\protect\ref{fig_z_a5.0_convergence:fig}(b).
The VMC error on the last digit is indicated in parenthesis.}
\label{tab_etot_a5.0:tab}
\end{table}

Our findings are in agreement with previous variational studies of
lattice models with short-range interactions
which showed that a correct description of the Mott-Hubbard insulating phase
can be achieved only by combining a Gutzwiller projector 
and a long-range Jastrow factor.
\cite{Capello05a,Miyagawa2011}

In fact, although the single H atoms are on average neutral,
charge fluctuations which give, e.g., virtual H$^+$-H$^-$ pairs,
are always possible.
Such charge fluctuations are in fact strongly suppressed at large 
interatomic distance, $a$ (and, in the Hubbard model, for large $U$).
Therefore, at large $a$, in order to prevent an instability of the Mott-Hubbard 
insulator toward a metallic state driven by the quantum fluctuations,
\cite{Mott49}
the H$^+$-H$^-$ pairs must be bound.
In the context of the Hubbard model,
\cite{Capello05a,Miyagawa2011}
it has been demonstrated that such binding mechanism can be included 
in the variational wave function by means of an inhomogeneous
Jastrow factor, analogous to the the second term in Eq.~\ref{jastrow}.
Crucially, the matrix element $g_{\alpha,\beta}$ in the expansion 
of the inhomogeneous Jastrow factor (See. Eq.~\ref{jastroweq}) can be
nonvanishing also for pairs of orbitals (labeled by the Greek indices)
which are far apart.
Indeed, such long range correlation is necessary to bind the 
virtual H$^+$-H$^-$ pairs and, 
along with the homogeneous Jastrow factor, provides an accurate
way to model electron localization in the Mott-Hubbard insulating phase,
as shown in Fig.~\ref{fig_z_a5.0_convergence:fig}(a).
%
\section{Discussion and conclusions}\label{conclusions}
%
In this article, we have investigated the ground state of a homogeneous, 
linear, and periodic chain of Hydrogen atoms (or H-chain) from first principles,
by means of a state-of-the-art variational Monte Carlo approach.
In fact, using a highly correlated 
Jastrow-antisymmetrized geminal power
variational wave function allowed us to obtain a total energy per atom 
comparable to benchmark density matrix renormalization group
calculations.
\cite{Hachmann06}
Furthermore, by using the so-called ``modern theory of polarization'',
\cite{Kohn64,*Resta94,*Hine07}
we have characterized the crossover between the weakly 
correlated (for small interatomic distances)
and the strongly correlated regimes (for large 
interatomic distances) through a single, simple  parameter, i.e., the
electronic localization length.
Our results extend to the properly periodic H-chain previous results
obtained by studying long, yet finite, chains.
\cite{Hachmann06,Tsuchimochi09,Sinitskiy10,*Mazziotti11}
In particular, we confirmed that the crossover
between the weakly and strongly correlated regimes of the H-chain 
corresponds physically to a crossover between 
a (finite-size) metallic and an insulating phase.
Finally, by studying the asymptotic behavior of the spin-spin 
and dimer-dimer correlation functions, we have also verified
that the insulating phase of the real H-chain is of the 
Mott-Hubbard type, as expected from the Hubbard model.
\cite{Hubbard64,*Gebhard_book}
Since the correct description of such a correlated insulator is
beyond the possibility of density functional theory in any of its
conventional local or semi-local approximations, one can think of
using our findings to devise an improved class of functionals.
We are currently exploring this possibility.

Intriguingly, we have found small but noticeable deviations from 
the behavior predicted by the Hubbard model
in the case of the scaling exponent of the dimer-dimer correlation 
function predicted by the Hubbard model.
\cite{Schulz90,*Fabrizio96,*Capello05b}
These deviations can be possibly due to the finite-size scaling or to
a true discrepancy between the H-chain and the 1D Hubbard model 
in the thermodynamic limit.

It will be interesting to check for possible new low energy physics 
of the H-chain at variance of the one-band, 
1D Hubbard model predictions.
These might be originated by:
i) The long-range Coulomb repulsion---indeed inefficiently screened 
in 1D systems---not included in the Hubbard model;
ii) The atomic orbital polarization---essential to describe 
non-covalent contribution to the bonding---not representable in terms 
of $1s$ orbitals, only.
In particular, relative simple elaborations of the Hubbard model, 
which just contain next-nearest neighbor interaction,
already predict a rich phase diagram even for a 1D system.
\cite{Capello05a,Lai10}
Besides, it is known that long-range Coulomb repulsion
can yield gapless charge excitations (plasmons) in 1D,
as observed in experiments.
\cite{Schulz93,*Goni91}     

Of course, more accurate finite-size extrapolation is desirable, 
although not possible with our current computational resources.
In particular, the use of diffusion Monte Carlo to improve 
the variational results might be also beneficial, but in practice 
still highly problematic due to the well known 
ergodicity issues in dealing with strong electronic localization 
in 1D systems.
\cite{Casula05,*Casula06}

In conclusion, given that the homogeneous, linear and periodic chain 
of Hydrogen atoms is becoming a standard test-case for \emph{ab initio}
approaches to strong electronic correlation,
\cite{Hachmann06,Tsuchimochi09,Sinitskiy10,*Mazziotti11}
the results reported in this article show that
variational Monte Carlo (with a highly correlated
Jastrow-antisymmetrized geminal power variational wave function) 
can provide an accurate and flexible
alternative to highly correlated methods of quantum chemistry.
Besides, and at variance with most of the methods of quantum chemistry,  
variational Monte Carlo can be successfully employed to study a strongly 
correlated solid in low dimensions close to a crossover 
or a phase transition.
\cite{Resta99,Capello05a}
%
\begin{acknowledgments}
This work has been performed under the HPC-EUROPA2 project (project number: HPC08TH6KD)
with the support of the European Commission - Capacities Area - Research Infrastructures
and using HPC resources from GENCI-IDRIS (project No.~100063). 
We acknowledge also financial support from the Spanish MEC (FIS2011-65702-C02-01), 
ACI-Promociona (ACI2009-1036), Grupos Consolidados UPV/EHU del Gobierno Vasco (IT-319-07),
and the European Research Council Advanced Grant DYNamo (ERC-2010-AdG-Proposal No. 267374).
Finally, we thanks Garnet K.-L. Chan, Simone Fratini, and Markus Holzmann  
for the help provided.
\end{acknowledgments}
%
%
%
\end{document}